\documentclass[10pt]{article}
\usepackage{amsmath,esdiff}
\usepackage{amssymb}
\usepackage{authblk}
\usepackage[normalem]{ulem}
\usepackage{bbold}
\usepackage{geometry}
\usepackage{setspace}
\usepackage{cite}
\usepackage[colorlinks=true,linkcolor=blue,citecolor=red,urlcolor=red]{hyperref}%
\usepackage{mathrsfs}
\usepackage{hyperref}
\usepackage{graphicx}
\usepackage{cancel}
\usepackage{array,esint}
\usepackage[most]{tcolorbox}
\graphicspath{}
\begin{document}
\title{\textbf{Probing the generalized uncertainty principle through quantum noises in optomechanical systems}}
\author{\textbf{$\mathbf{Soham}$ $\mathbf{Sen}^{a\dagger}$, $\mathbf{Sukanta}$ $\mathbf{Bhattacharyya}^{b*}$ and $\mathbf{Sunandan}$ $\mathbf{Gangopadhyay}^{a\ddagger}$
\footnote{{\quad}\\
{${}^\dagger$ soham.sen@bose.res.in, sensohomhary@gmail.com}\\
{${}^*$sukanta706@gmail.com}\\
{${}^\ddagger$sunandan.gangopadhyay@bose.res.in, sunandan.gangopadhyay@gmail.com}}}}
\affil{{${}^a$ Department of Theoretical Sciences}\\
{S.N. Bose National Centre for Basic Sciences}\\
{JD Block, Sector III, Salt Lake, Kolkata 700 106, India}\\
{${}^b$ Department of Physics}\\ {West Bengal State University, Barasat, Kolkata 700 126, India}}
\date{}
\maketitle
\begin{abstract}
\noindent In this work we have considered a simple mechanical oscillator interacting with a single mode optical field inside a cavity in the generalized uncertainty principle framework (GUP). Our aim is to calculate the modified noise spectrum and observe the effects of the GUP. The commutation relation that we have considered has an extra linear order momentum term along with a quadratic order term. Confronting our theoretical results with the observational results, we observe that we get a much tighter bound on the GUP parameters from the noise spectrum using the values of the system parameters from different experiments.
\end{abstract}
\section{Introduction}
The two most revolutionary theories of the past century are general theory of relativity and quantum mechanics. Classical gravity described by Einstein's field equations is one of the most accurate theories of the universe describing very precise fundamental aspects of spacetime. On the other hand we know that the fundamental building blocks of our universe follow quantum mechanical laws. Therefore the search for the Planck-scale nature of gravity is very important. Loop quantum gravity \cite{ROVELLI,CARLIP}, noncommutative geometry \cite{GLO} and string theory \cite{ACV, KOPAPR} have been able to give a theoretical framework for a quantum theory of gravity. All the extensive studies in these field suggests that there should be an observer independent minimal length (the Planck length $\approx 10^{-33}$ m). The minimal length can be incorporated by modifying the standard Heisenberg Uncertainty Principle, which we call as the generalized uncertainty principle or GUP. The relation between the minimal length and gravity was shown first in \cite{BRON1, BRON2} and later in \cite{MEAD}. Along with the theoretical models there is a very strong signature of the existence of the GUP in the gedanken experiments as well. To find the existence of GUP there have been several investigations in black hole physics \cite{MAGGIORE, SCARDIGLI, ADLERSAN, ADLERCHENSAN, RABIN, SG1, SCARDIGLI2, SG2, Ong, EPJC, BMajumder}, cosmology \cite{Giardino}, quantum gravity corrections in simple harmonic oscillator \cite{DAS1, DAS2}, path integral framework of a particle in the GUP formalism moving under a potential \cite{DAS3, SG3} and so on. Recently there have been several efforts to devise a laboratory test to search for effects of GUP  in optomechanical experiments\cite{Petruzzeillo, DasModak}.  Incorporating GUP into a theory also introduces some parameters with unknown bounds, also known as GUP parameters and the determination of the order of such parameters will help to understand the effect of GUP in the theoretical realm. Various investigations have been done to find an upper bound to the GUP parameter $\gamma_0$ \cite{SCARDIGLI2, DAS1, BM, FENG, BBGK, SCARDIGLI3, SG4, RCSG}. Similar calculation with the deformed commutation relations has been used to explain quantum optical phenomena \cite{IVASP, IVASP2, KSP}.
In case of an optomechanical system one can search for indirect signatures of quantum gravity in low energy tabletop experiments. It was first shown in \cite{IVASP} that if there is a controlled interaction between an optical field and a mechanical oscillator then the output optical field may depend on the mechanical commutation relation and if the measurements of the average value of the field are done to the highest precision level, it will be possible to test GUP. In \cite{GRD} a sensitive tabletop experiment is proposed where the effect of changing the commutator affects macroscopic oscillator motion in a noise bath experiment. They found new bounds for the optomechanical parameter by comparing the noise spectrum result to the experimental data sets available. 

\noindent In this paper once again we have considered a mechanical oscillator interacting with a single mode optical field inside a cavity. Further, we have taken a modified commutation relation incorporating linear and quadratic order momentum corrections and with this set up, we carry out the analysis of the noise spectrum. The modified momentum in \cite{SGSB} has been used to calculate bounds on the GUP parameters. We have then used our form of the modified noise spectrum and used parameters from several tabletop optomechanical and gravitational wave experiments to calculate the final bounds on the GUP parameters. We observe that the bounds on the GUP parameters improved significantly due to consideration of a linear order correction in the uncertainty relation.
\section{Noise spectrum in the generalized uncertainty principle framework}
\noindent Our aim is to determine the noise spectrum for a mechanical oscillator due to thermal and radiation pressure noise in the presence of the generalized uncertainty principle (GUP) incorporating both linear and quadratic order contributions in momentum uncertainty. To begin our analysis, we start by writing down the modified uncertainty relation between the position $\tilde{x}_i$ and its conjugate momentum $\tilde{p}_j$ given as follows \cite{Farag, Farag2, SGSB}
\begin{equation}\label{1.0}
\Delta \tilde{x}_i\Delta\tilde{p}_j\geq \frac{\hbar}{2}\left[1-\alpha\left\langle \tilde{p}+\frac{\tilde{p}_i\tilde{p}_j}{\tilde{p}}\right\rangle-\left(\alpha^2-\gamma\right)\left((\Delta \tilde{p})^2+\langle\tilde{p}\rangle^2\right)-\left(\alpha^2-2\gamma\right) \left((\Delta \tilde{p}_i)^2+\langle\tilde{p}_i\rangle^2\right)\right]~.
\end{equation}
The form of the modified commutation relation satisfied by $\tilde{x}_i$ and $\tilde{p}_j$ consistent with eq.(\ref{1.0}) reads
\begin{equation}\label{2.0}
[\tilde{x}_i,\tilde{p}_j]=i\hbar\left[\delta_{ij}-\alpha\left(\delta_{ij}\tilde{p}+\frac{\tilde{p}_i\tilde{p}_j}{\tilde{p}}\right)+\gamma(\delta_{ij}\tilde{p}^2+2\tilde{p}_i\tilde{p}_j)-\alpha^2 (\delta_{ij}\tilde{p}^2+\tilde{p}_i\tilde{p}_j)\right]~.
\end{equation}
For our case $i, j=1$ since we shall work in one dimension, therefore we will not be using any indices explicitly. In this case the modified commutation relation reads,
\begin{equation}\label{1.1}
[\tilde{x},\tilde{p}]=i\hbar\left[1-2\alpha \tilde{p}-2\alpha^2\tilde{p}^2+3\gamma\tilde{p}^2\right]~.
\end{equation}
 The modified variables $(\tilde{x},\tilde{p})$ in terms of the usual variables $(x,p)$ following from the commutation relation in eq.(\ref{1.1})\cite{Farag, Farag2, SGSB} reads
\begin{equation}\label{2.1}
\tilde{x}= x~, ~~~\tilde{p}=p-\alpha p^2+\gamma p^3~.
\end{equation}
\noindent In this paper we have used the modified momentum representation to incorporate the effect of GUP in our analysis. $p$ in eq.(\ref{2.1}) has the standard coordinate representation $-i\hbar\frac{\partial}{\partial x}$. Therefore, in the coordinate representation the modified momentum $\tilde{p}$ has the following representation
\begin{equation}\label{1.2}
\tilde{p}=-i\hbar\frac{\partial}{\partial x}+\alpha\hbar^2\frac{\partial^2}{\partial x^2}-i\gamma\hbar^3\frac{\partial^3}{\partial x^3}~.
\end{equation}

\noindent The system that we are interested in involves a typical optomechanical system \cite{ASP} consisting of a mechanical oscillator of mass `$m$' inside a cavity interacting with an optical field. The mechanical oscillator is coupled to a thermal bath at temperature $T$ with damping rate $\rho$. Then the Hamiltonian describing the system in presence of the GUP reads
\begin{equation}\label{2.2}
H=\frac{\tilde{p}^2}{2m}+\frac{1}{2}m\Omega^2 \tilde{x}^2+\hbar \omega_c a^{\dagger} a-\hbar G \tilde{x} a^\dagger a~.
\end{equation}
\noindent Here $\omega_c$ is the angular frequency of the optical cavity and $\Omega$ is the vibrational frequency of the oscillator. The coupling constant of the mechanical oscillator and the optical field is defined as $G=\frac{\omega_c}{L}$, where $L$  is the length of the cavity. It is noted that $a$ and $a^{\dagger}$ are respectively the annihilation and creation operators of the optical field. Note that in principle we could have incorporated GUP in the optical field also by deforming the commutation relation between $a$ and $a^\dagger$\cite{BossoMann}. However, as we shall see below that this would make our analysis very complicated. Hence, for simplicity we have kept the standard commutation relation between $a$ and $a^\dagger$. Also, because of the $\hbar$ in the last two terms of eq.(\ref{2.2}), the modification of $a$ and $a^\dagger$ by GUP would make these terms significantly smaller than the corrections considered in the present case. Using the relation (\ref{2.1}), the Hamiltonian (\ref{2.2}) can be recast in terms of the usual variables $(x,p)$ upto order $\mathcal{O} (\alpha^2, \gamma)$ \cite{SGSB}.
\begin{align}
H&\simeq\frac{1}{2}m\Omega^2 x^2-\hbar G x a^\dagger a+\hbar \omega_c a^{\dagger} a+\frac{p^2}{2m}-\frac{\alpha p^3}{m}+\frac{(\alpha^2+2\gamma) p^4}{2m}\label{2.3}
\end{align} 
\noindent where $x$ and $p$ are the position and momentum operators of the mechanical oscillator following the usual commutation relation, $[x,p]=i\hbar$. We can break the Hamiltonian (\ref{2.3}) as 
\begin{equation}
H = H_0 + V_{mod}	
\label{2.4}
\end{equation}
where 
\begin{align}
H_0& = \frac{1}{2}m\Omega^2 x^2-\hbar G a^\dagger a+\hbar \omega_c a^{\dagger} a+\frac{p^2}{2m}~,\label{2.5}\\
V_{mod}&= -\frac{\alpha p^3}{m}+\frac{(\alpha^2+2\gamma) p^4}{2m}~.\label{2.6}
\end{align}
\noindent The above expressions show that $H_0 $ is the usual Hamiltonian describing the optomechanical system. The effect of the GUP incorporating both the linear and quadratic order contributions in momentum uncertainty actually appears through the presence of $V_{mod}$.
In this paper we will follow the linearization treatment carried out in \cite{GRD, ASP}. To do this, we write down the optical field around the mean field $\mathcal{A}$ as
\begin{equation}\label{2.7}
a=\mathcal{A}+\delta a
\end{equation}
\noindent where $\langle a\rangle= \mathcal{A}$ is the average coherent amplitude, which depends on the drive power $P$ and the optical decay rate $\kappa$. Here $\delta a$ is a small fluctuation term . Therefore, we keep only the terms linear in $\delta a $ and $(\delta a)^{\dagger} $.  We also linearize the position and momentum operators about $x_0$ and $p_0$, where $x_0$ and $p_0$ are the position and momentum operators in the absence of any GUP corrections.  
Hence, we can write the position and momentum operators as \footnote[1]{Note that throughout this paper we will assume that the cavity is driven on resonance.} follows 
\begin{align}
x&\simeq x_0+\delta x\label{2.8}\\
p&\simeq p_0+\delta p\label{2.9}~.
\end{align}

\noindent Therefore, the Langevin equations of mechanics  upto $\mathcal{O}(\alpha^2,~\gamma)$ are given as follows
\begin{align}
\dot{x}&=\dot{x_0}+\delta\dot{x}\simeq \frac{p_0+\delta p}{m}-\frac{3\alpha p_0^2}{m}+\frac{2(\alpha^2+2\gamma)p_0^3}{m}\label{2.10}\\
\dot{p}&=\dot{p_0}+\delta\dot{p}\simeq -m\Omega^2(x_0+\delta x)+f-\rho m (\dot{x}_0+\delta\dot{x})~.\label{2.11}
\end{align}
\noindent In eq.(\ref{2.11}), $\rho m (\dot{x}_0+\delta\dot{x})$ indicates the viscous damping factor \cite{GRDZOL} and $f$ consists of a stochastic thermal force $f_T$ along with $f_D$, where $f_D$ contains the radiation pressure force $f_{rad}$ and contributions due to other forces on the oscillator. We will keep only $f_{rad}$ as our main source of force as other contributions will be negligible for our working frequency range. For understanding $f$, we  need to simplify the term with the optomechanical coupling constant using the linearization of `$a$' in eq.(\ref{2.7})
\begin{equation}\label{2.12}
\begin{split}
\hbar G a^\dagger a =& \hbar G (\mathcal{A}^*+\delta a^{\dagger})(\mathcal{A}+\delta a)~.
\end{split}
\end{equation} 
\noindent Considering $\mathcal{A}$ to be real without any loss of generality, and keeping up to linear order in $\delta a $ and $\delta a^\dagger$, we identify the operator corresponding to the radiation pressure as follows
\begin{equation}\label{2.13}
f_{rad}=\hbar G \mathcal{A}(\delta a+\delta a^\dagger)~.
\end{equation}
The stochastic thermal force and the radiation pressure force follows the following relations for the correlation functions \cite{GRD}
\begin{align}
&\langle f_T(\sigma_1) f_T(\sigma_2)\rangle=2 k_B T\rho m \delta(\sigma_1-\sigma_2)\label{2.14}~,\\
&\langle f_{rad}(\sigma_1)f_{rad}(\sigma_2)\rangle=\hbar^2 G^2 \mathcal{A}^2 e^{-\frac{\kappa\left|\sigma_1-\sigma_2\right|}{2}}\label{2.15}~.
\end{align}
The derivation of the relation (\ref{2.15}) is given in Appendix A1.

\noindent Now for $x_0$ and $p_0$, the equations of motion can be recast in the matrix form as
\begin{align}
\begin{pmatrix}
\dot{x}_0\\
\dot{p}_0
\end{pmatrix}&=
\begin{pmatrix}
0&\frac{1}{m}\\
-m \Omega^2&-\rho
\end{pmatrix}
\begin{pmatrix}
x_0\\
p_0
\end{pmatrix}+\begin{pmatrix}
0\\
f
\end{pmatrix}\label{2.16}
\end{align}
\noindent and the perturbative corrections are
\begin{align}
\begin{pmatrix}
\dot{\delta x}\\
\dot{\delta p}
\end{pmatrix}&=
\begin{pmatrix}
0&\frac{1}{m}\\
-m \Omega^2&-\rho
\end{pmatrix}
\begin{pmatrix}
\delta x\\
\delta p
\end{pmatrix}+\begin{pmatrix}
-\frac{3\alpha p_0^2}{m}+\frac{2(\alpha^2+2\gamma)p_0^3}{m}\\
3\alpha\rho p_0^2-2(\alpha^2+2\gamma)\rho p_0^3
\end{pmatrix}~.
\label{2.17}
\end{align}

\noindent After getting the equations of motion, we can immediately find their solutions. Eq.(\ref{2.16}) actually describes a damped driven oscillator with frequency $\Omega$ and damping rate $\rho$. Therefore, for underdamped regime $\Omega>\frac{\rho}{2}$, the eigenvalues of the system are $\lambda_{\pm}=-\rho_0\pm i\omega_0\quad$ with $\rho_0=\frac{\rho}{2}$ and $\omega_0=\frac{\sqrt{4\Omega^2-\rho^2}}{2}$. The solutions for the unperturbed position and momentum are 
\begin{align}
x_0(t)&=\frac{1}{2\omega_0}\left[i e^{\lambda_+ t}\left(x_0 \lambda_--\frac{p_0}{m}\right)-i e^{\lambda_- t}\left(x_0 \lambda_+-\frac{p_0}{m}\right)+\frac{i \xi}{m}-\frac{i \xi^*}{m}\right]\label{2.18}~,\\
p_0(t)&=\frac{i}{2\omega_0}\left[m\lambda_+e^{\lambda_+ t}\left(x_0 \lambda_--\frac{p_0}{m}\right)-m\lambda_-e^{\lambda_- t}\left(x_0 \lambda_+-\frac{p_0}{m}\right)+\frac{\lambda_-\xi}{m}-\frac{\lambda_+\xi^*}{m}\right]\label{2.19}~,\\
\xi(t)&=\int_{0}^{t}f(\sigma)e^{\lambda_-(t-\sigma)}d\sigma~,\label{2.20}
\end{align} 
\noindent and the solutions for the perturbed position and momentum are given as follows
\begin{align} 
\delta x(t)&=\frac{1}{2\omega_0}\left[i e^{\lambda_+ t}\left(\delta x(0) \lambda_--\frac{\delta p(0)}{m}\right)+\frac{i \chi}{m}+\frac{i \zeta}{m}+h.c.\right]\label{2.21}~,\\
\delta p(t)&=\frac{1}{2\omega_0}\left[i \lambda_+e^{\lambda_+ t}\left(\delta x(0) \lambda_-m-\delta p(0)\right)-i \lambda_+\chi^*-i \lambda_+\zeta^*+h.c.\right]\label{2.22}~,
\end{align}
where $\chi$ and $\zeta$ read 
\begin{equation}
\chi(t)=-3\lambda_-\alpha\int_{0}^{t} e^{\lambda_-(t-\sigma)}p_0^2\hspace{1mm} d\sigma~,~\zeta(t)=2\lambda_-(\alpha^2+2\gamma)\int_{0}^{t} e^{\lambda_-(t-\sigma)}p_0^3 \hspace{1mm} d\sigma~.
\label{2.23}\end{equation}

\noindent In this paper our basic goal is to study the power spectrum due to position fluctuation in the steady state of the mechanical oscillator in presence of the GUP. The power spectrum is defined as \cite{ASP}
\begin{equation}\label{2.24}
S_{PF}(\omega)=\int_{-\infty}^{\infty}\left\langle x(\tau)x(0)\right\rangle e^{i\omega\tau}d\tau~.
\end{equation}
\noindent We are basically doing a quantum mechanical analysis, therefore $S_{PF}(\omega) \neq S_{PF}(-\omega)$. Hence, we study the symmetrised noise spectrum $S_{PF}(\omega) =[S_{PF}(\omega)+S_{PF}(-\omega)]/2$.

\noindent Setting $\delta x(0)=\delta p(0)=0$ at $t=0$, we can write the correlation function of the positions at two different times upto order $\mathcal{O}( \alpha^2, \gamma)$ as
\begin{equation}\label{2.25}
\begin{split}
\left\langle x(\tau)x(0)\right\rangle&=\left\langle (x_0(\tau)+\delta x(\tau))(x_0(0)+\delta x(0))\right\rangle\simeq \left\langle (x_0(\tau)+\delta x(\tau))x_0(0)\right\rangle\\&=\left\langle x_0(\tau)x_0(0)\right\rangle+\left\langle \delta x(\tau)x_0(0)\right\rangle~.
\end{split}
\end{equation}
\noindent We will now calculate the above correlation function in order to obtain the power spectrum of the mechanical oscillator. To obtain the first term in the RHS of eq.(\ref{2.25}), we use eq.(\ref{A2.6}) in Appendix A2 and substitute $x(\tau)x(0)$ in place of $\hat{A}$. This leads to the following expression
\begin{equation}\label{2.26}
\begin{split}
\langle x_0(\tau)x_0(0)\rangle_H&\simeq\langle x_0(\tau)x_0(0)\rangle_{H_0}-\langle x_0(\tau)x_0(0)\rangle_{H_0} \left\langle\int_{0}^{\beta}d\beta'\int_{0}^{\beta'}d\beta''  e^{\beta' H_0}V_1e^{-\beta' H_0}  e^{\beta'' H_0}V_1e^{-\beta'' H_0}\right\rangle_{H_0}
\\&+\langle x_0(\tau)x_0(0)\rangle_{H_0} \left\langle \int_0^\beta d\beta' e^{\beta' H_0}V_2 e^{-\beta' H_0}\right\rangle_{H_0}-\left\langle \int_0^\beta d\beta' e^{\beta' H_0}V_2 e^{-\beta' H_0}x_0(\tau)x_0(0)\right\rangle_{H_0}
\end{split}
\end{equation}
\begin{equation*}
+\left\langle\int_{0}^{\beta}d\beta'\int_{0}^{\beta'}d\beta''  e^{\beta' H_0}V_1e^{-\beta' H_0}  e^{\beta'' H_0}V_1e^{-\beta'' H_0}x_0(\tau)x_0(0)\right\rangle_{H_0}~.
\end{equation*}
with
\begin{equation}\label{2.27}
V_1=-\frac{\alpha p^3}{m}\approx-\frac{\alpha p_0^3}{m} ~,~ V_2=\frac{(\alpha^2+2\gamma) p^4}{2m}\approx \frac{(\alpha^2+2\gamma) p_0^4}{2m}~.
\end{equation}
Inserting the solution of $x_0(t)$ from eq.(\ref{2.16}) in $\langle x_0(\tau)x_0(0)\rangle_{H_0}$, and using the notations $x_0(0)=x_0$~and~$p_0(0)=p_0$, we obtain\footnote{The suffix R denotes the real part of the correlation. Here we will be using the real part of the results only as the noise spectrum is an observable quantity.}
\begin{equation}
\langle x_0(\tau)x_0\rangle_{H_0,R}=\frac{\langle x_0^2\rangle_{H_0}}{2\omega_0}\left(i\lambda_-e^{\lambda_+\tau}-i\lambda_+e^{\lambda_-\tau}\right)\label{2.28}~,
\end{equation}
\begin{equation}
\langle\xi x_0\rangle_{H_0,R}=\langle x_0p_0\rangle_{H_0,R}=0\nonumber~. 
\end{equation}  

\noindent The position and momentum operators of the quantum mechanical oscillator are
\begin{align}
p_0&=i P_\mathcal{O}(b^\dagger-b)\label{2.29}~,\\
x_0&=X_\mathcal{O}(b+b^\dagger)\label{2.30}~,\\
P_\mathcal{O}&=\sqrt{\frac{m\hbar\Omega}{2}}~,~X_\mathcal{O}=\sqrt{\frac{\hbar}{2 m\Omega}}\nonumber~.
\end{align}
In eq.(s)(\ref{2.29}, \ref{2.30}), $b$ and $b^\dagger$ follows the usual commutation relation. Using the forms of $p_0$ and $x_0$ in eq.(s)(\ref{2.29}, \ref{2.30}) we can calculate the following relations
\begin{align}
&e^{\beta \hat{H}_0}p_0~e^{-\beta \hat{H}_0}=iP_\mathcal{O}(b^\dagger e^{\beta \hbar\Omega}-b\hspace{1mm} e^{-\beta \hbar\Omega})\label{2.31}\\
&x_0^2=X_\mathcal{O}^2(b^2+{b^\dagger}^2+2 b^\dagger b+1)\label{2.32}~.
\end{align} 
Note that, every term consisting of equal powers of $b$ and $b^\dagger$ contributes to the expectation value. In our calculation we will use the relation $\langle {b^\dagger}^k{b}^k\rangle=k! \langle b^\dagger b\rangle^k$ \cite{TDJ}. We have done all the remaining calculations in the high temperature limit that is, $\beta \hbar\Omega\ll 1$. We have used the data from \cite{MRTY, MRTYth, pur, teu} to compute the bounds on the GUP parameters from the modified noise spectrum. In these experiments $\beta\hbar\Omega$ is of the order of ($10^{-3}-10^{-13}$) justifying our choice of taking the $\beta\hbar\Omega\ll 1$ limit. Using eq.(s)(\ref{2.31}, \ref{2.32}) in the high temperature limit, we arrive at the following results 
\begin{align}
&\left\langle \int_{0}^{\beta}d\beta'\int_{0}^{\beta'} d\beta''e^{\beta'H_0}V_1e^{-\beta' H_0}e^{\beta''H_0}V_1e^{-\beta''H_0}\right\rangle_{H_0}\approx\frac{\beta^2}{2}\frac{15 \alpha^2{\langle x_0^2\rangle}^3 ~P_{\mathcal{O}}^6}{m^2 ~ X_{\mathcal{O}}^6}\label{2.33},\\
&\left\langle \int_{0}^{\beta}d\beta'\int_{0}^{\beta'} d\beta''e^{\beta'H_0}V_1e^{-\beta'H_0}e^{\beta''H_0}V_1e^{-\beta''H_0}x_0^2\right\rangle_{H_0}\approx\frac{15\alpha^2 P_{\mathcal{O}}^6\beta^2}{2 m^2}\left(\frac{\langle x_0^2\rangle^4}{X_{\mathcal{O}}^6}-6\frac{\langle x_0^2\rangle^2}{X_{\mathcal{O}}^2}\right)\label{2.34},\\
&\left\langle \int_{0}^{\beta}d\beta'\int_{0}^{\beta'} d\beta''e^{\beta'H_0}V_2e^{-\beta'H_0}\right\rangle_{H_0}=3\beta P_{\mathcal{O}}^4\frac{\langle x_0^2\rangle^2}{X_{\mathcal{O}}^4}\label{2.35},\\
&\left\langle \int_{0}^{\beta}d\beta'\int_{0}^{\beta'} d\beta''e^{\beta'H_0}V_2e^{-\beta'H_0}x_0^2\right\rangle_{H_0}\approx 3\beta P_{\mathcal{O}}^4\left(\frac{\langle x_0^2\rangle^3}{X_{\mathcal{O}}^4}-4\langle x_0^2\rangle\right)\label{2.36}~.
\end{align}
Using the above results we obtain
\begin{equation}\label{2.37}
\begin{split}
\left\langle x_0(\tau)x_0(0)\right\rangle_{H,R}\approx&\left\langle x_0(\tau)x_0(0)\right\rangle_{H_0,R}+\left(\frac{3}{2}\left(\frac{\alpha^2}{2}+\gamma\right)\frac{\hbar^2}{\omega_0}\frac{m\Omega^2}{k_BT'}\left\langle x_0^2\right\rangle-\frac{45\alpha^2}{8}\frac{\hbar^2}{\omega_0}\frac{m^2\Omega^4}{k_B^2{T'}^2}\left\langle x_0^2\right\rangle^2\right)(i\lambda_-e^{\lambda_+\tau}-i\lambda_+e^{\lambda_-\tau})~.
\end{split}
\end{equation}
The above result gives the first term in the right hand side of eq.(\ref{2.25}). We now proceed to calculate the second term on the right hand side of eq.(\ref{2.25}). To do this we use the form of $\delta x(\tau)$ from eq.(\ref{2.21}). This gives

\begin{equation}\label{2.38}
\left\langle \delta x(\tau) x_0\right\rangle_{H,R} =\int_0^\tau (i \lambda_-e^{\lambda_-(\tau-\sigma)}-i \lambda_+e^{\lambda_+(\tau-\sigma)})\left(-\frac{3\alpha}{2\omega_0 m}\langle p_0(\sigma)^2x_0\rangle+\frac{\alpha^2+2\gamma}{m\omega_0}\langle p_0(\sigma)^3 x_0\rangle \right)d\sigma~.
\end{equation}
In the right hand side of eq.(\ref{2.38}), the first term in the parentheses will have no contribution. The exact expression of the second term, $\left\langle \delta x(\tau)x_0(0)\right\rangle$ in eq.(\ref{2.38}) is given by eq.(\ref{A3.6}) in Appendix A3.  Now using eq.(s)(\ref{2.37}, \ref{A3.6}) in eq.(\ref{2.25}), we can get the correlation function of the positions at different times over the modified Hamiltonian (\ref{2.3}) in presence of the GUP corrections, containing both the linear and quadratic terms in momentum uncertainty. In this paper we want to investigate the effect of GUP in the power spectrum of an optomechanical system. After getting the correlation function over the modified Hamiltonian (\ref{2.3}) we can now calculate the power spectrum $S_{PF}(\omega)$ using eq.(\ref{2.24}). To do this, we substitute eq.(s)(\ref{2.37}, \ref{2.38}) in eq.(\ref{2.24}), to get
\begin{equation}\label{2.39}
\begin{split}
S_{PF}(\omega)&=\int_{-\infty}^{\infty}\langle x(\tau)x(0)\rangle_{H,R} e^{i\omega\tau}d\tau\\
&=\int_{-\infty}^{\infty}\left(\langle x(\tau)x(0)\rangle_{H,R}+\langle \delta x(\tau)x(0)\rangle_{H,R}\right) e^{i\omega\tau}d\tau\\
&=\int_{-\infty}^{\infty}\left(\langle x(\tau)x(0)\rangle_{H_0,R}+\delta\langle x(\tau)x(0)\rangle_{H_0,R}\right) e^{i\omega\tau}d\tau\\
&=S_0(\omega)+\delta S(\omega)~.
\end{split}
\end{equation}
The form of $S_0(\omega)$ in the right hand side of eq.(\ref{2.39}) is given by \cite{ASP, Clerk} 
\begin{equation}\label{2.40}
\begin{split}
	S_0(\omega)=&\int_{-\infty}^{\infty}\langle x(\tau)x(0)\rangle_{H_0,R}~ e^{i\omega\tau}d\tau\\
	=&\frac{\kappa^2+4\omega^2}{16\kappa\mathcal{A}^2G^2}+\frac{4\kappa\hbar^2\mathcal{A}^2G^2}{(\kappa^2+4\omega^2)m^2(\rho^2\omega^2+(\omega^2-\Omega^2)^2)}+\frac{2\rho k_BT}{m(\rho^2\omega^2+(\omega^2-\Omega^2)^2)}~.
\end{split}
\end{equation}

\noindent The form of the perturbed noise spectrum, showing the effect of GUP upto order $\mathcal O(\alpha^2,\gamma)$ can be calculated as follows. 

\noindent The second term in the right hand side of eq.(\ref{2.37}) can be simplified as
\begin{equation}\label{2.41}
\begin{split}
&\left(\frac{3}{2}\left(\frac{\alpha^2}{2}+\gamma\right)\frac{\hbar^2}{\omega_0}\frac{m\Omega^2}{k_BT'}\left\langle x_0^2\right\rangle-\frac{45\alpha^2}{8}\frac{\hbar^2}{\omega_0}\left(\frac{m\Omega^2}{k_BT'}\right)^2\left\langle x_0^2\right\rangle^2\right)(i\lambda_-e^{\lambda_+\tau}-i\lambda_+e^{\lambda_-\tau})\\
&=\frac{3}{2}\frac{\rho_0^2+\omega_0^2}{\omega_0^2}\left(\gamma-\frac{13}{4}\alpha^2\right)\left\langle x_0^2\right\rangle \left(\mathcal{P}_1+i\mathcal{Q}_1\right)e^{(-\rho_0-i\omega_0)\tau}+c.c~,\\
&\mathcal{P}_1=\frac{\hbar^2m\omega_0^2}{k_BT'}~,~\mathcal{Q}_1=\frac{\hbar^2m\rho_0\omega_0}{k_BT'}~.
\end{split}
\end{equation}
Combining eq.(\ref{2.41}) with eq.(\ref{A3.6}) (see Appendix A3), we obtain the result for $\delta \langle x(\tau)x(0)\rangle_{H_0, R}$ as 
\begin{equation}\label{2.42}
\begin{split}
\delta\left\langle x(\tau) x(0)\right\rangle_{H_0,R}&=\frac{3}{2}(\alpha^2+2\gamma)\left\langle x_0^2\right\rangle \frac{\rho_0^2+\omega_0^2}{\omega_0^2}\biggr\{e^{(-3\rho_0+i3\omega_0)\tau}(a_1+i f_1)+e^{(-3\rho_0+i\omega_0)\tau}(a_2+i f_2)\\&+e^{(-\rho_0-i\omega_0)\tau}(p_2+i q_2)-\tau e^{(-\rho_0-i\omega_0)\tau}(u_1+i v_1)+e^{\left(-2\rho_0+2i\omega_0-\frac{\kappa}{2}\right)\tau}(u_2+iv_2)\\&+e^{\left(-2\rho_0-\frac{\kappa}{2}\right)\tau}\mathcal{K}\biggr\}+\frac{3}{2}\left(\gamma-\frac{13}{4}\alpha^2\right)\left\langle x_0^2\right\rangle \frac{\rho_0^2+\omega_0^2}{\omega_0^2}\left(\mathcal{P}_1+i\mathcal{Q}_1\right)e^{(-\rho_0-i\omega_0)\tau}+c.c.
\end{split}
\end{equation}
with the exact forms of $a_1, f_1, a_2, f_2, p_2, q_2, u_1, v_1, \mathcal{P}_1, \mathcal{Q}_1, u_2, v_2, \mathcal{K}$ given respectively in eq.(s)(\ref{A5.8}-\ref{A5.14}) of Appendix A5. The Fourier transform of eq.(\ref{2.42}) is the perturbed spectrum, which is given as follows (We keep in mind that the entire expression is valid for $\tau\geq0$ and in the regime $\tau< 0$ it has no contribution)
\begin{equation}\label{2.43}
\begin{split}
\delta S(\omega)&=\int_0^\infty \delta\langle x(\tau) x(0)\rangle_{H_0,R}~e^{i\omega\tau}d\tau\\
&\simeq\frac{6k_BT'}{m\omega_0^2}\left(\frac{\alpha^2}{2}+\gamma\right)\biggr\{\frac{2\mathcal{K}\kappa}{\kappa^2+4\omega^2}+\frac{q_2\omega_0(\rho_0^2-\omega^2+\omega_0^2)+p_2\rho_0(\rho_0^2+\omega^2+\omega_0^2)}{\omega^4+2\omega^2(\rho_0^2-\omega_0^2)+(\rho_0^2+\omega_0^2)^2}\\
&+\frac{3 a_2\rho_0(9\rho_0^2+\omega^2+\omega_0^2)-f_2\omega_0(9\rho_0^2-\omega^2+\omega_0^2)}{\omega^4+2\omega^2(9\rho_0^2-\omega_0^2)+(9\rho_0^2+\omega_0^2)^2}\\&
+\frac{3 a_1\rho_0(9\rho_0^2+\omega^2+9\omega_0^2)+3f_1\omega_0(9\rho_0^2-\omega^2+9\omega_0^2)}{\omega^4+18\omega^2(\rho_0^2-\omega_0^2)+81(\rho_0^2+\omega_0^2)^2}\\
&-\{-u_1\omega^6+(\rho_0^2+\omega_0^2)^2(u_1(\rho_0^2-\omega_0^2)+2v_1\rho_0\omega_0)-\omega^4(u_1(\rho_0^2-\omega_0^2)+6v_1\rho_0\omega_0)\\&+\omega^2(u_1(\rho_0^4+10\rho_0^2\omega_0^2+\omega_0^4)-4v_1\rho_0\omega_0(\rho_0^2-\omega_0^2))\}/\{\omega^4+2\omega^2(\rho_0^2-\omega_0^2)+(\rho_0^2+\omega_0^2)^2\}\\
&+\frac{-2v_2\omega_0\left(\left(\frac{\kappa}{2}+2\rho_0\right)^2-(\omega^2-4\omega_0^2)^2\right)+u_2\left(\frac{\kappa}{2}+2\rho_0\right)\left(\left(\frac{\kappa}{2}+2\rho_0\right)^2+(\omega^2+4\omega_0^2)^2\right)}{\left(\left(\frac{\kappa}{2}+2\rho_0\right)^2+(\omega^2+4\omega_0^2)^2\right)^2-16\omega^2\omega_0^2}
\biggr\}\\&+\frac{3k_BT'}{m\omega_0^2}\left(-\frac{13\alpha^2}{4}+\gamma\right)\frac{\mathcal{P}_1\rho_0(\rho_0^2+\omega^2+\omega_0^2)+\mathcal{Q}_1\omega_0(\rho_0^2-\omega^2+\omega_0^2)}{\omega^4+2\omega^2(\rho_0^2-\omega_0^2)+(\rho_0^2+\omega_0^2)^2}~.
\end{split}
\end{equation}
Here, $k_BT'$ is related to $k_BT$ via the following relation
\begin{equation}\label{2.44}
k_BT'=k_BT+\frac{8h\nu\mathcal{F}^2P}{\pi^2 c^2\rho m}
\end{equation}
where $\mathcal{F}=\frac{\pi c}{\kappa L}$ is the fineness of the cavity.
\noindent Eq.(\ref{2.43}) is one of the main findings in this paper. The following plot is an overall representation of the modified spectrum with respect to the unperturbed noise spectrum.

\begin{figure}[h]
\includegraphics[scale=0.3]{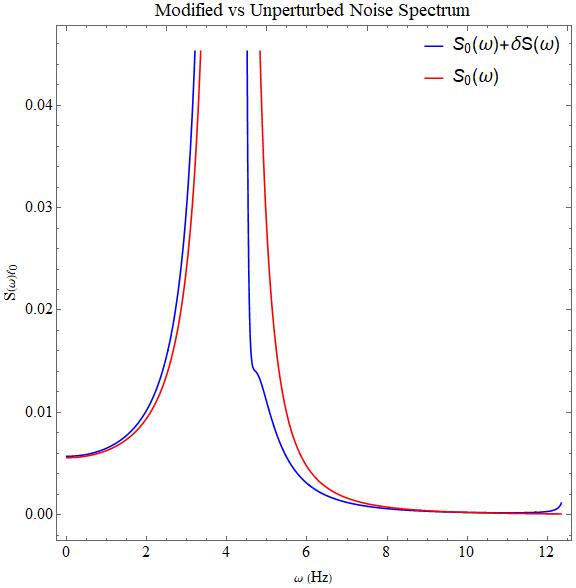}
\centering
\caption{Plot of $S_0(\omega)+\delta S(\omega) \text{ and } S_0(\omega) \text{ vs }\omega$ using the parameters in Table 1}
\end{figure}
\noindent In Figure 1, $\ell_{0}$ is a scaling factor which has been used to amplify the details of the individual noise spectrum. In Figure 1, the correction to the noise spectrum ($\delta S(\omega)$) has been amplified in order to make the effects of the perturbation more visible in the plot. We observe that the modified noise spectrum have higher values compared to the unperturbed noise spectrum just below the resonance frequency (for $\omega\leq4.15$ Hz) and lower values above it. The modified spectrum again attains higher values after $\omega\geq10$ Hz.

\section{Obtaining the Bounds on the GUP parameters}
In this section we will proceed to calculate the bounds on the GUP parameters $\alpha$ and $\gamma$. Now let us consider some optomechanical experiments to obtain the bounds on the GUP parameters $\alpha$ and $\gamma$. First we consider the experimental result from advanced laser interferometer gravitational-wave observatory (aLIGO). 
\begin{table}
\centering
\caption{Experimental parameters for aLIGO gravitational wave experiment}
	\medskip
	\begin{tabular}{c||c}
		\textbf{Experimental Parameters}& \textbf{aLIGO} \cite{MRTY,MRTYth}\\
		\hline\hline
		$T$ $(K)$&$3.00\times 10^2$\\
		$\Omega$ $(Hz)$& $4.15$\\
		$\rho$ $(Hz)$&$1\times 10^{-6}$\\
		$Q$&$1.33\times 10^9$\\
		$\nu$ $(Hz)$&$2.82\times 10^{14}$\\
		$L(m)$& $4.00\times 10^3$\\
		$\kappa$ $(Hz)$& $4.78\times 10^3$\\
		$m$ $(kg)$& $10$\\
		$P$ $(W)$&$3.60\times 10^3$\\
		$\mathcal{F}$&$4.92\times 10^1$\\
		$S(\omega)$ $(Minimum)$ $(m^2/Hz)$&$9.00\times 10^{-40}$
	\end{tabular}
\end{table}
In aLIGO there are two long perpendicular arms (4 km each) and through the arms a very strong laser beam is reflected back. Hence the interference is observed mainly to detect the gravitational waves via shifts in the light intensity. Recent observed total noise spectrum \cite{MRTY, MRTYth} in aLIGO shows a gap of order one magnitude between observed and expected values of the noise spectrum in the frequency range $20$-Hz to $100$-Hz. Within this frequency range thermal noise dominants over shot noise. Therefore we can directly use the values of the parameters in the form of the perturbed noise spectrum in eq.(\ref{2.43}) bearing the effects of GUP. Now the parameters of a single-oscillator within a single-cavity in resonance ($\omega=\Omega$) for aLIGO, as we consider in this paper, has been given in Table 1. Note that the relation $\delta S(\omega) \leq S_{0}$ will help us to obtain bounds on the GUP parameters $\alpha$ and $\gamma$. Using the value of the parameters from Table 1 in eq.(\ref{2.43}) we get the relations for $\delta S(\omega)$ and $S_0(\omega)$ as (we will be writing the form without mentioning units)
\begin{align}
\delta S(4.15)&\cong(8.007\alpha^2+16.014\gamma)\times 10^{-37}~(m^2/Hz)~\label{3.0a},\\
S_0(4.15)&\cong9.621\times 10^{-17}~(m^2/Hz)\label{3.0b}~.
\end{align}
 To get the final bound we used the inequality $\gamma\geq 3.5 \alpha^2$ from \cite{SGSB} in eq.(\ref{3.0a}) and compared it with eq.(\ref{3.0b}) from which we can directly obtain the bound for $\alpha$. For comparing the bounds we will use $\alpha_0$ and $\gamma_0$ instead of $\alpha$ and $\gamma$, where the relations between $\alpha_0$ and $\alpha$, and $\gamma_0$ and $\gamma$ are given as
\begin{align}
\alpha&=\frac{\alpha_0}{m_pc}\label{3.1}~,\\
\gamma&=\frac{\gamma_0}{m_p^2c^2}\label{3.2}~,\\
 m_p=\text{Planck}~&\text{mass}~,~c=\text{Speed of light}~.
\end{align}
Using relations (\ref{3.1}, \ref{3.2}), the final bounds on the modified GUP parameters are obtained as, $\alpha_0\leq 10^{10}$ and $\gamma_0\leq 10^{20}$. The reason for using $\alpha_0$ and $\gamma_0$ is that they are dimensionless parameters.

\noindent In another scenario \cite{pur}, a silicon nitride membrane oscillator is placed in a Fabry-Perot cavity. At resonance the oscillator is driven by the radiation pressure fluctuations of a laser beam. In this case the radiation pressure noise of power spectrum can be measured. Again in \cite{teu}, an LC circuit is used as a microwave cavity. A coherent microwave drive at the cavity resonance is applied via a feed-line. In this case the radiation fluctuations dominates the thermal noise. Therefore for the both cases contain a single optomechanical cavity with a mechanical oscillator, and so we can insert the values of the parameters from Table 2 directly in eq.(\ref{2.43}) to obtain the bounds on the GUP parameters for the above experiments.
The values of the parameters for the above experiments are given in Table 2.
\begin{table}
\centering
\caption{Experimental parameters for tabletop optomechanical experiments}
\medskip
\begin{tabular}{c||c||c}
\textbf{Experimental Parameters}& \textbf{Purdy \textit{et} \textit{al}} \cite{pur}& \textbf{Teufel \textit{et} \textit{al}} \cite{teu}\\
\hline\hline
$T$ $(K)$&$1.70\times 10^{-3}$&$4.00\times 10^{-2}$\\
$\Omega$ $(Hz)$& $9.75\times 10^6$& $5.88\times 10^7$\\
$\rho$ $(Hz)$&$8.98\times 10^{3}$&$1.53\times 10^{2}$\\
$Q$&$1.08\times 10^3$&$3.83\times 10^5$\\
$\nu$ $(Hz)$&$2.82\times 10^{14}$&$6.71\times 10^{9}$\\
$L(m)$& $5.10\times 10^{-3}$& $4.00\times 10^{-8}$\\
$\kappa$ $(Hz)$& $5.59\times 10^6$& $6.64\times 10^7$\\
$m$ $(kg)$& $7\times 10^{-12}$& $8.50\times 10^{-14}$\\
$P$ $(W)$&$9.40\times 10^{-5}$&$7.80\times 10^{-9}$\\
$\mathcal{F}$&$3.30\times 10^4$&$3.55\times 10^8$\\
$S(\omega)$ $(Minimum)$ $(m^2/Hz)$&$4.40\times 10^{-32}$&$1.00\times 10^{-26}$\\
$\gamma_0=$ $upper$ $bound$ $@$ $\omega=\Omega$&$1\times 10^{41}$&$1\times 10^{42}$
\end{tabular}
\end{table}
\noindent Substituting the parameters from the second data column of Table 2 in eq.(\ref{2.43}) and calculating $S_0(\omega)$ and $\delta S(\omega)$ as earlier, we get the bounds as $\alpha_0\lesssim 10^{19} $ and $\gamma_0\lesssim 10^{38}$ for experiment \cite{pur} at the resonance frequency.  In a similar manner we get the bounds $\alpha_0\lesssim 10^{19}$ and $\gamma_0\lesssim 10^{38}$ for experiment \cite{teu} at the resonance frequency. From our findings, we observe that the bounds on the GUP parameters are much tighter than those obtained earlier in the literature \cite{DAS1, SG3, SG4, RCSG, GRD}.
\section{Conclusion}
In this work, we have considered a typical optomechanical set up consisting of a simple harmonic oscillator interacting with a field in a cavity. The entire system is analysed in the framework of the generalized uncertainty principle. In our analysis, we have considered both the linear and quadratic order terms in momentum uncertainty in  contrast to the earlier works which considered the simplest form of the generalized uncertainty principle. The main part of our investigation involved the computation of the noise spectrum from the two point correlation function of the position observables of the oscillator in the high temperature limit and under the resonance condition. The theoretical result obtained consists of the usual noise spectrum corrected by a piece coming from the generalized uncertainty principle parameters. Demanding that this correction must be smoother than the usual spectrum, we obtain bounds on both the linear as well as quadratic GUP parameters. We find that the bounds are much tighter than those obtained earlier in the literature in different physical scenarios.
\pagebreak
\section*{Appendix}
\subsection*{A1. Calculating the steady state expectation value of the radiation pressure force}
In this appendix we will calculate the steady state expectation value of the radiation pressure force by beginning from the following equations \cite{ASP}
\begin{equation}\label{A1.1}
\hbar G\mathcal{A}\hspace{0.5 mm}\delta \dot{a}=-\frac{\kappa}{2}\hbar G\mathcal{A}\delta a+\hbar G\mathcal{A}\sqrt{\kappa}f_{\mathcal{C}}~,
\end{equation}
\begin{equation}\label{A1.2}
\hbar G\mathcal{A}\hspace{0.5mm}\delta \dot{a}^{\dagger}=-\frac{\kappa}{2}\hbar G\mathcal{A}\delta a^\dagger+\hbar G\mathcal{A}\sqrt{\kappa}f_{\mathcal{C}}^\dagger~.
\end{equation}
Combining eq.(s)(\ref{A1.1}, \ref{A1.2}) we get:
\begin{equation}\label{A1.3}
\begin{split}
\hbar G\mathcal{A}(\delta \dot{a}+\delta \dot{a}^{\dagger})&=-\frac{\kappa}{2}\hbar G\mathcal{A}(\delta a+\delta a^\dagger)+\hbar G\mathcal{A}\sqrt{\kappa}(f_{\mathcal{C}}+f_{\mathcal{C}}^\dagger)\\
\implies \dot{f}_{rad}&=-\frac{\kappa}{2}f_{rad}+\hbar G\mathcal{A}\sqrt{\kappa}(f_{\mathcal{C}}+f_{\mathcal{C}}^\dagger)~.
\end{split}
\end{equation}
The solution to eq.(\ref{A1.3}) can be calculated as follows:
\begin{equation}\label{A1.4}
f_{rad}(\tau)=f_{rad}(0)e^{-\frac{\kappa\tau}{2}}+\sqrt{\kappa}\hbar G\mathcal{A}\int\limits_0^\tau d\mathcal{T}e^{-\frac{\kappa(\tau-\mathcal{T})}{2}}(f_{\mathcal{C}}(\mathcal{T})+f_{\mathcal{C}}^\dagger(\mathcal{T}))~.
\end{equation}
The stochastic term $f_{\mathcal{C}}$ is assumed to have white noise correlation as
\begin{equation}\label{A1.5}
\langle f_{\mathcal{C}}(\tau_1) f_{\mathcal{C}}(\tau_2)\rangle= \delta(\tau_1-\tau_2)~.
\end{equation}
This then leads to the steady state correlation function between $f_{rad}(\tau_1)$ and $f_{rad}(\tau_2)$. This reads
\begin{equation}\label{A1.6}
\langle f_{rad}(\tau_1)f_{rad}(\tau_2)\rangle=\hbar^2 G^2 \mathcal{A}^2 e^{-\frac{\kappa\left|\tau_1-\tau_2\right|}{2}}~.
\end{equation}
\subsection*{A2. Thermal perturbation theory}
In this Appendix our goal is to calculate the thermal average $\langle \hat{A}\rangle_H$ of an observable $\hat{A}$, where $\hat{H}=\hat{H}_0+\hat{V}_{mod}$. 
To calculate the aforementioned expectation value we define an operator as follows
\begin{equation}\label{A2.1}
\hat{\mu}(\beta)=e^{\beta \hat{H}_0}e^{-\beta \hat{H}}~.
\end{equation}
Taking derivative with respect to $\beta$ of the above eq.(\ref{A2.1}) we get the following differential equation involving $\hat{\mu}(\beta)$
\begin{equation}\label{A2.2}
\frac{d\hat{\mu}(\beta)}{d\beta}=-e^{\beta \hat{H}_0}~\hat{V}_{mod}~e^{-\beta \hat{H}_0}\hat{\mu}(\beta)~.
\end{equation}
We shall obtain the solution upto second order in $\hat{V}_{mod}$ since we are considering terms upto second order in $\alpha$ and first order in $\gamma$. The solution of eq.(\ref{A2.2}) upto second order in $\hat{V}_{mod}$ is given by 
\begin{equation}\label{A2.3}
\hat{\mu}(\beta)\simeq 1-\int_{0}^{\beta}d\beta'  e^{\beta' \hat{H}_0}~\hat{V}_{mod}~e^{-\beta' \hat{H}_0}+\int_{0}^{\beta}d\beta'  e^{\beta' \hat{H}_0}~\hat{V}_{mod}~e^{-\beta' \hat{H}_0}\int_{0}^{\beta'}d\beta''  e^{\beta'' \hat{H}_0}~\hat{V}_{mod}~e^{-\beta'' \hat{H}_0}~.
\end{equation}
The thermal average of an observable $A$ upto second order in $\hat{V}_{mod}$ reads
\begin{align}
\langle \hat{A}\rangle_H=&Tr(Z_H^{-1} e^{-\beta \hat{H}}\hat{A})\label{A2.4}\\
=&Z_0Z_H^{-1}\left(1-\left\langle \int_{0}^{\beta}d\beta'  e^{\beta' \hat{H}_0}\hat{V}_{mod}e^{-\beta' \hat{H}_0}\hat{A}\right\rangle_{H_{0}}+\left\langle\int_{0}^{\beta}d\beta'  \int_{0}^{\beta'}d\beta''e^{\beta' \hat{H}_0}\hat{V}_{mod}e^{-\beta' \hat{H}_0} e^{\beta'' \hat{H}_0}\hat{V}_{mod}e^{-\beta'' \hat{H}_0}\hat{A}\right\rangle_{H_0}\right)\nonumber~.
\end{align}
Now putting $\hat{A}$=$\mathbb{1}$ in the above expression, we get
\begin{equation}\label{A2.5}
\begin{split}
Z_0 Z_H^{-1}\cong &~ 1+\left\langle \int_{0}^{\beta}d\beta'  e^{\beta' \hat{H}_0}~\hat{V}_{mod}~e^{-\beta' \hat{H}_0}\right\rangle_{H_{0}}+\left\langle \int_{0}^{\beta}d\beta'  e^{\beta' \hat{H}_0}~\hat{V}_{mod}~e^{-\beta' \hat{H}_0}\right\rangle_{H_{0}}\left\langle \int_{0}^{\beta}d\beta''  e^{\beta'' \hat{H}_0}~\hat{V}_{mod}~e^{-\beta'' \hat{H}_0}\right\rangle_{H_{0}}\\
&-\left\langle\int_{0}^{\beta}d\beta' \int_{0}^{\beta'}d\beta'' e^{\beta' \hat{H}_0}~\hat{V}_{mod}~e^{-\beta' \hat{H}_0} e^{\beta'' \hat{H}_0}~\hat{V}_{mod}~e^{-\beta'' \hat{H}_0}\right\rangle_{H_0}.
\end{split}
\end{equation}
Substituting the form of $Z_0Z_H^{-1}$ in eq.(\ref{A2.4}) and keeping terms up to second order in $\hat{V}_{mod}$, we get the expression for $\langle\hat{A}\rangle_H$ as
\begin{equation}\label{A2.6}
\begin{split}
\langle \hat{A}\rangle_H&=\langle A\rangle_{H_0}+\langle \hat{A}\rangle_{H_0}\left\langle \int_{0}^{\beta}d\beta'  e^{\beta' \hat{H}_0}\hat{V}_{mod}e^{-\beta' \hat{H}_0}\right\rangle_{H_{0}}\left\langle \int_{0}^{\beta}d\beta''  e^{\beta'' \hat{H}_0}\hat{V}_{mod}e^{-\beta'' \hat{H}_0}\right\rangle_{H_{0}}\\
&-\langle \hat{A}\rangle_{H_0} \left\langle\int_{0}^{\beta}d\beta'\int_{0}^{\beta'}d\beta''  e^{\beta' \hat{H}_0}\hat{V}_{mod}e^{-\beta' \hat{H}_0}  e^{\beta'' \hat{H}_0}\hat{V}_{mod}e^{-\beta'' \hat{H}_0}\right\rangle_{H_0}-\left\langle \int_{0}^{\beta}d\beta'  e^{\beta' \hat{H}_0}\hat{V}_{mod}e^{-\beta' \hat{H}_0}\hat{A}\right\rangle_{H_{0}}\\
&+\langle \hat{A}\rangle_{H_0}\left\langle \int_{0}^{\beta}d\beta'  e^{\beta' \hat{H}_0}\hat{V}_{mod}e^{-\beta' \hat{H}_0}\right\rangle_{H_{0}}+\left\langle\int_{0}^{\beta}d\beta'\int_{0}^{\beta'}d\beta''  e^{\beta' \hat{H}_0}\hat{V}_{mod}e^{-\beta' \hat{H}_0}  e^{\beta'' \hat{H}_0}\hat{V}_{mod}e^{-\beta'' \hat{H}_0}\hat{A}\right\rangle_{H_0}\\
&-\left\langle \int_{0}^{\beta}d\beta'  e^{\beta' \hat{H}_0}\hat{V}_{mod}e^{-\beta' \hat{H}_0}\right\rangle_{H_{0}}\left\langle \int_{0}^{\beta}d\beta''  e^{\beta'' \hat{H}_0}\hat{V}_{mod}e^{-\beta'' \hat{H}_0}\hat{A}\right\rangle_{H_{0}}~.
\end{split}
\end{equation}
We can simplify eq.(\ref{A2.6}) using the relations 
\begin{align}
\frac{Z_1}{Z_0}&=-\left\langle \int_{0}^{\beta}d\beta'  e^{\beta' \hat{H}_0}~\hat{V}_{mod}~e^{-\beta' \hat{H}_0}\right\rangle_{H_{0}}\label{A2.7}
\end{align}
and
\begin{align}
\frac{Z_2}{Z_0}&=\frac{\beta}{2}\int_{0}^{\beta}d\beta'\left\langle  e^{\beta' \hat{H}_0}~\hat{V}_{mod}~e^{-\beta' \hat{H}_0}~\hat{V}_{mod}\right\rangle_{H_{0}}\label{A2.8}.
\end{align}
After some simple steps we arrive at the final form of the thermal average of $\hat{A}$
\begin{equation}\label{A2.9}
\begin{split}
\langle\hat{A}\rangle_H&=\langle \hat{A}\rangle_{H_0}-\frac{Z_1}{Z_0}\langle \hat{A}\rangle_{H_0}+\left(\frac{Z_1}{Z_0}\right)^2\langle \hat{A}\rangle_{H_0}-\left(1-\frac{Z_1}{Z_0}\right)\left\langle \int_0^\beta d\beta' e^{\beta' H_0}~\hat{V}_{mod}~e^{-\beta' H_0}\hat{A}\right\rangle_{H_0}\\
&-\frac{Z_2}{Z_0}\langle \hat{A}\rangle_{H_0}+\left\langle \int_{0}^{\beta}d\beta'\int_{0}^{\beta'}d\beta''e^{\beta' H_0}~\hat{V}_{mod}~e^{(\beta''-\beta') H_0}~\hat{V}_{mod}~ e^{-\beta'' H_0}\hat{A}\right\rangle_{H_0}~.
\end{split}
\end{equation}
This is the form which we shall use in the main text. We shall also drop the hat symbol on the operators for convenience.
\subsection*{A3. Results for the perturbation term}
In this Appendix we will calculate the $\left\langle\delta x(\tau)x(0)\right\rangle_{H}$ term. Using eq.(\ref{2.19}), $p_0$ in eq.(\ref{2.38}) can be recast in the following way 
\begin{align}
p_0(\sigma)&=u(\sigma)p_0+v(\sigma)x_0+q(\sigma)\label{A3.1}
\end{align}
where 
\begin{align}
u(\sigma)&=\frac{1}{2\omega_0}(i\lambda_-e^{\lambda_-\sigma}-i\lambda_+e^{\lambda_+\sigma})\label{A3.2}~,\\
v(\sigma)&=\frac{\lambda_+\lambda_-m}{2\omega_0}(i e^{\lambda_-\sigma}-i e^{\lambda_+\sigma})\label{A3.3}~,\\
q(\sigma)&=\frac{1}{2\omega_0}(i\lambda_-\xi(\sigma)-i\lambda_+\xi^*(\sigma))~\label{A3.4}~.
\end{align}
Using eq.(\ref{A3.1}), eq.(\ref{2.38}) can be rewritten in the following form
\begin{equation}\label{A3.5}
\begin{split}
\left\langle\delta x(\tau) x_0\right\rangle_{H}=\frac{\alpha^2+2\gamma}{m\omega_0}\int_0^\tau(i\lambda_-e^{\lambda_-\sigma}-i\lambda_+e^{\lambda_+\sigma})
\left(v_\sigma^3\left\langle x_0^4\right\rangle_{H_0}+3 v_\sigma\left\langle q_\sigma^2\right\rangle_{H_0}\left\langle x_0^2\right\rangle_{H_0}+3u_\sigma^2v_\sigma\left\langle x_0^2 p_0^2\right\rangle_{H_0}\right)d\sigma\end{split}
\end{equation}
where $v(\sigma)=v_\sigma$, $u(\sigma)=u_\sigma$ and $q(\sigma)=q_\sigma$.
It is easy to find that $\left\langle x_0^4\right\rangle_{H_0}=3\left\langle x_0^2\right\rangle^2_{H_0}$ and $\left\langle x_0^2 p_0^2\right\rangle_{H_0}\approx \left\langle x_0^2\right\rangle_{H_0} \left\langle p_0^2\right\rangle_{H_0}$.
Here we will use correlations (\ref{2.14}, \ref{2.15}) to reach the final solution after several steps. The result obtained is as follows
\begin{equation}\label{A3.6}
\begin{split}
\left\langle \delta x(\tau) x_0\right\rangle_{H_0}=&-6(\alpha^2+2\gamma)\frac{\lambda_+\lambda_-}{(\lambda_+-\lambda_-)^2}\left\langle  x_0^2\right\rangle_{H_0}\biggr[e^{3\lambda_+\tau}\biggr\{\left\langle  x_0^2\right\rangle_{H_0}\frac{-3(\lambda_+\lambda_-m)^2 }{2(3\lambda_+-\lambda_-)(\lambda_--\lambda_+)}\\&+\left\langle  p_0^2\right\rangle_{H_0}\frac{-3\lambda_+^2}{2(3\lambda_+-\lambda_-)(\lambda_--\lambda_+)}+2k_BT\rho m\frac{-3\lambda_+}{4(3\lambda_+-\lambda_-)(\lambda_--\lambda_+)}\\&+\frac{-3\hbar^2G^2\mathcal{A}^2\lambda_+}{(3\lambda_+-\lambda_-)(\lambda_--\lambda_+)(\kappa+2\lambda_+)}\biggr\}+e^{(2\lambda_++\lambda_-)\tau}\biggr\{\left\langle  x_0^2\right\rangle_{H_0}\frac{3(\lambda_+\lambda_-m)^2 (2\lambda_++\lambda_-)}{2\lambda_+(\lambda_++\lambda_-)(\lambda_--\lambda_+)}\\&+\left\langle  p_0^2\right\rangle_{H_0}\frac{(2\lambda_++\lambda_-)(\lambda_++2\lambda_-)}{2(\lambda_++\lambda_-)(\lambda_--\lambda_+)}+2k_BT\rho m\frac{(2\lambda_++\lambda_-)(\lambda_++5\lambda_-)}{4(\lambda_++\lambda_-)^2(\lambda_--\lambda_+)}\\&
+\frac{\hbar^2 G^2\mathcal{A}^2(2\lambda_++\lambda_-)(6\lambda_-(\lambda_++\lambda_-)+\kappa(\lambda_++5\lambda_-))}{(\kappa+2\lambda_-)(\lambda_--\lambda_+)(\lambda_++\lambda_-)^2(\kappa+2\lambda_+)}\biggr\}\\&
+e^{\lambda_-\tau}\biggr\{\left\langle  x_0^2\right\rangle_{H_0}\frac{3(\lambda_--\lambda_+)(\lambda_+\lambda_-m)^2}{2\lambda_+(3\lambda_+-\lambda_-)(\lambda_++\lambda_-)}+\left\langle  p_0^2\right\rangle_{H_0}\frac{\lambda_-(\lambda_--\lambda_+)}{2(\lambda_++\lambda_-)(3\lambda_+-\lambda_-)}\\&
+2k_BT\rho m\frac{3\lambda_+^3+7\lambda_-\lambda_+^2-3\lambda_-^2\lambda_++\lambda_-^3}{2(3\lambda_+-\lambda_-)(\lambda_++\lambda_-)^2(\lambda_--\lambda_+)}\\&+2\hbar^2G^2\mathcal{A}^2\biggr\{-24\lambda_-^2(\lambda_--\lambda_+)^2\lambda_+(\lambda_-^2-\lambda_-\lambda_+-2\lambda_+^2)-\kappa^4(\lambda_-^3-3\lambda_-^2\lambda_++7\lambda_-\lambda_+^2\\&+3\lambda_+^3)-2\kappa^2\lambda_+(25\lambda_-^4-41\lambda_-^3\lambda_+-5\lambda_-^2\lambda_+^2+57\lambda_-\lambda_+^3+12\lambda_+^4)+\kappa^3(5\lambda_-^4+\lambda_-^3\lambda_+\\&-23\lambda_-^2\lambda_+^2+47\lambda_-\lambda_+^3+18\lambda_+^4)+4\kappa\lambda_-(\lambda_-^5+\lambda_-^4\lambda_++11\lambda_-^3\lambda_+^2-35\lambda_-^2\lambda_+^3+20\lambda_-\lambda_+^4\\&+18\lambda_+^5)\biggr\}\biggr/\biggr((\kappa-2\lambda_-)^2(\kappa-4\lambda_++2\lambda_-)(\lambda_--3\lambda_+)(\kappa-2\lambda_+)^2(\lambda_--\lambda_+)(\lambda_++\lambda_-)^2\biggr)
\\&\biggr\}-\tau e^{\lambda_-\tau}\biggr\{2k_BT\rho m\frac{\lambda_-}{2(\lambda_++\lambda_-)}+\frac{2\hbar^2\mathcal{A}^2G^2\kappa\lambda_-}{(\kappa-2\lambda_-)(\kappa-2\lambda_+)(\lambda_++\lambda_-)}\biggr\}\\&
-e^{(2\lambda_+-\frac{\kappa}{2})\tau}\frac{16\hbar^2\mathcal{A}^2G^2\kappa\lambda_+(\kappa-4\lambda_+)}{(\kappa+2\lambda_-)(\kappa-4\lambda_++2\lambda_-)(\kappa-2\lambda_+)^2(\kappa+2\lambda_+)}\\&
+e^{(\lambda_++\lambda_--\frac{\kappa}{2})\tau}\frac{16\hbar^2\mathcal{A}^2G^2\kappa\lambda_+(\kappa-2\lambda_+-2\lambda_-)}{(\kappa-2\lambda_+)^2(\kappa+2\lambda_+)(\kappa^2-4\lambda_-^2)}+c.c\biggr]~.
\end{split} 
\end{equation}
\subsection*{A4. Some essential expectation values}
In this section we will record few expectation values. These are essential for calculating the final form of the noise spectrum : 
\begin{equation}\label{A4.1}
\left\langle x_0^2\right\rangle_{H_0}=\lim_{\substack{t\rightarrow\infty}}\left\langle \left(\frac{1}{2m\omega_0}
\left(i\xi(t)-i\xi^*(t)\right)\right)^2\right\rangle=\frac{\hbar^2\mathcal{A}^2G^2(\kappa+4\rho_0)}{\rho_0m^2\Omega^2((\kappa+2\rho_0)^2+4\omega_0^2)}~,
\end{equation}
\begin{equation}\label{A4.2}
\left\langle p_0^2\right\rangle_{H_0}=\lim_{\substack{t\rightarrow\infty}}\left\langle \left(\frac{1}{2\omega_0}
\left(i\lambda_-\xi(t)-i\lambda_+\xi^*(t)\right)\right)^2\right\rangle=\frac{\hbar^2\mathcal{A}^2G^2\kappa}{\rho_0((\kappa+2\rho_0)^2+4\omega_0^2)}~.
\end{equation}
For optomechanical experiments we can assume that $\kappa\gg \rho$. This assumption helps us to consider that due to the radiative fluctuation and the thermal bath, the unperturbed optomechanical oscillator is in a steady state with an effective temperature of $T'$. 
Using this approximation we find from eq.(s)(\ref{A4.1}, \ref{A4.2}) that
\begin{align}
\frac{\left\langle p_0^2\right\rangle_{H_0}}{2m}&=\frac{m\Omega^2\left\langle x_0^2\right\rangle_{H_0}}{2}\quad\{\textit{when }\kappa\gg \rho\}~,\label{A4.3}\\
\frac{m\Omega^2\left\langle x_0^2\right\rangle_{H_0}}{2}&=\frac{k_BT'}{2}=\frac{k_BT}{2}+ \frac{\hbar^2\mathcal{A}^2G^2\kappa}{2\rho_0m(\kappa^2+4\omega_0^2)}~.\label{A4.4}
 \end{align}
\subsection*{A5. The exact and generalized form of the perturbed spectrum}
In this section we will calculate the exact and approximate form of the perturbed spectrum using the expectation values in Appendix A4. We denote the following terms in the beginning in order to explicitly write the exact form of the perturbed correlation function :
\begin{align}
A1=&
\left\langle  x_0^2\right\rangle_{H_0}\frac{-3(\lambda_+\lambda_-m)^2 }{2(3\lambda_+-\lambda_-)(\lambda_--\lambda_+)}+\left\langle  p_0^2\right\rangle_{H_0}\frac{-3\lambda_+^2}{2(3\lambda_+-\lambda_-)(\lambda_--\lambda_+)}\nonumber\\&+2k_BT\rho m\frac{-3\lambda_+}{4(3\lambda_+-\lambda_-)(\lambda_--\lambda_+)}+\frac{-3\hbar^2G^2\mathcal{A}^2\lambda_+}{(3\lambda_+-\lambda_-)(\lambda_--\lambda_+)(\kappa+2\lambda_+)}\label{A5.1}~,\\
A2=&\left\langle  x_0^2\right\rangle_{H_0}\frac{3(\lambda_+\lambda_-m)^2 (2\lambda_++\lambda_-)}{2\lambda_+(\lambda_++\lambda_-)(\lambda_--\lambda_+)}+\left\langle  p_0^2\right\rangle_{H_0}\frac{(2\lambda_++\lambda_-)(\lambda_++2\lambda_-)}{2(\lambda_++\lambda_-)(\lambda_--\lambda_+)}\nonumber\\&+2k_BT\rho m\frac{(2\lambda_++\lambda_-)(\lambda_++5\lambda_-)}{4(\lambda_++\lambda_-)^2(\lambda_--\lambda_+)}+\frac{\hbar^2 G^2\mathcal{A}^2(2\lambda_++\lambda_-)(6\lambda_-(\lambda_++\lambda_-)+\kappa(\lambda_++5\lambda_-))}{(\kappa+2\lambda_-)(\lambda_--\lambda_+)(\lambda_++\lambda_-)^2(\kappa+2\lambda_+)}\label{A5.2}~,\\
B1=&\left\langle  x_0^2\right\rangle_{H_0}\frac{3(\lambda_--\lambda_+)(\lambda_+\lambda_-m)^2}{2\lambda_+(3\lambda_+-\lambda_-)(\lambda_++\lambda_-)}+\left\langle  p_0^2\right\rangle_{H_0}\frac{\lambda_-(\lambda_--\lambda_+)}{2(\lambda_++\lambda_-)(3\lambda_+-\lambda_-)}\nonumber\\&
+2k_BT\rho m\frac{3\lambda_+^3+7\lambda_-\lambda_+^2-3\lambda_-^2\lambda_++\lambda_-^3}{2(3\lambda_+-\lambda_-)(\lambda_++\lambda_-)^2(\lambda_--\lambda_+)}\nonumber\\&+2\hbar^2G^2\mathcal{A}^2\biggr\{-24\lambda_-^2(\lambda_--\lambda_+)^2\lambda_+(\lambda_-^2-\lambda_-\lambda_+-2\lambda_+^2)-\kappa^4(\lambda_-^3-3\lambda_-^2\lambda_++7\lambda_-\lambda_+^2\nonumber\\&+3\lambda_+^3)-2\kappa^2\lambda_+(25\lambda_-^4-41\lambda_-^3\lambda_+-5\lambda_-^2\lambda_+^2+57\lambda_-\lambda_+^3+12\lambda_+^4)+\kappa^3(5\lambda_-^4+\lambda_-^3\lambda_+\nonumber\\&-23\lambda_-^2\lambda_+^2+47\lambda_-\lambda_+^3+18\lambda_+^4)+4\kappa\lambda_-(\lambda_-^5+\lambda_-^4\lambda_++11\lambda_-^3\lambda_+^2-35\lambda_-^2\lambda_+^3+20\lambda_-\lambda_+^4\nonumber\\&+18\lambda_+^5)\biggr\}\biggr/\biggr((\kappa-2\lambda_-)^2(\kappa-4\lambda_++2\lambda_-)(\lambda_--3\lambda_+)(\kappa-2\lambda_+)^2(\lambda_--\lambda_+)(\lambda_++\lambda_-)^2\biggr)\label{A5.3}~,\\
B2=&\frac{16\hbar^2\mathcal{A}^2G^2\kappa\lambda_+(\kappa-4\lambda_+)}{(\kappa+2\lambda_-)(\kappa-4\lambda_++2\lambda_-)(\kappa-2\lambda_+)^2(\kappa+2\lambda_+)}\label{A5.4}~,\\
C1=&\frac{16\hbar^2\mathcal{A}^2G^2\kappa\lambda_+(\kappa-2\lambda_+-2\lambda_-)}{(\kappa-2\lambda_+)^2(\kappa+2\lambda_+)(\kappa^2-4\lambda_-^2)}\label{A5.5}~.
\end{align}
The forms of $a_1, f_1, a_2, f_2, p_2, q_2, u_1, v_1, \mathcal{P}_1, \mathcal{Q}_1, u_2, v_2, \mathcal{K}$ in eq. (\ref{2.42}) in terms of the above parameters are given as
\begin{align}
a_1=\Re[A1]\approx-\frac{3\hbar^2\mathcal{A}^2G^2\omega_0^2}{(\kappa^2+4\omega_0^2)(\rho_0^2+4\omega_0^2)}
&~,~ f_1=\Im[A1]\approx-\frac{3\hbar^2\mathcal{A}^2G^2\kappa\omega_0}{4(\kappa^2+4\omega_0^2)(\rho_0^2+4\omega_0^2)}\label{A5.8}~,\\
a_2=\Re[A2]\approx-\frac{3\hbar^2\mathcal{A}^2G^2\kappa}{2\rho_0(\kappa^2+4\omega_0^2)}&~,~f_2=\Im[A2]\approx\frac{3\hbar^2\mathcal{A}^2G^2\kappa\omega_0}{4\rho_0^2(\kappa^2+4\omega_0^2)}\label{A5.9}~,\\
p_2=\Re[B1]\approx\frac{39\hbar^2\mathcal{A}^2G^2\omega_0^2\kappa^4}{(\rho_0^2+4\omega_0^2)(\kappa^2+4\omega_0^2)^2(\kappa^2+36\omega_0^2)}&~,~q_2=\Im[B1]\approx \frac{k_BT'm\rho_0}{\omega_0}-\frac{8\hbar^2\mathcal{A}^2G^2\omega_0\kappa(5\kappa^2+18\omega_0^2)(\rho_0^2+3\omega_0^2)}{(\rho_0^2+4\omega_0^2)(\kappa^2+4\omega_0^2)^2(\kappa^2+36\omega_0^2)}~,\label{A5.10}\\
u_1=k_BT'\rho_0m&~,~v_1=k_BT'\omega_0m~,\label{A5.11}\\
\mathcal{P}_1=\frac{\hbar^2 m\omega_0^2}{k_B T'}&~,~\mathcal{Q}_1=\frac{\hbar^2 m\rho_0\omega_0}{k_B T'}~,\label{A5.12}\\
u2=\Re[B2]\approx \frac{16\hbar^2\mathcal{A}^2G^2\kappa(\rho_0\kappa^4+6\kappa^3\omega_0^2+88\kappa\omega_0^4)}{(\kappa^2+4\omega_0^2)^3(\kappa^2+36\omega_0^2)}&~,~v2=\Im[B2]\approx-\frac{16\hbar^2\mathcal{A}^2G^2\kappa\omega_0(\kappa^4+12\kappa^2\omega_0^2-96\omega_0^4)}{(\kappa^2+4\omega_0^2)^3(\kappa^2+36\omega_0^2)}~,\label{A5.13}\\
\mathcal{K}=\Re[C1]&\approx-\frac{16\hbar^2\mathcal{A}^2G^2\kappa^2(\rho_0\kappa+2\omega_0^2)}{(\kappa^2+4\omega_0^2)^3}\label{A5.14}~.
\end{align}
Here the approximated forms are calculated in the $\kappa\gg\rho$ limit. In the white noise regime  (i.e. radiation pressure noise is similar to white noise), $\kappa\gg\Omega,\rho$. In this approximation we get $a_1=f_1=a_2=f_2=p_2=u_2=v_2=\mathcal{K}\cong 0$ and $q_2=\frac{k_BT'm\rho_0}{\omega_0}$.
Therefore, eq.(\ref{2.42}) reduces to the following form 
\begin{equation}\label{A5.15}
\begin{split}
\delta\left\langle x(\tau) x(0)\right\rangle_{H_0,R}&=\frac{3}{2}(\alpha^2+2\gamma)\left\langle x_0^2\right\rangle \frac{\rho_0^2+\omega_0^2}{\omega_0^2}\left[\left(i\frac{k_BT'm\rho_0}{\omega_0}-\tau mk_BT'(\rho_0+i\omega_0)\right)e^{(-\rho_0-i\omega_0)\tau}+c.c.\right]\\
&+\frac{3}{2}\left(\gamma-\frac{13}{4}\alpha^2\right)\left\langle x_0^2\right\rangle\frac{\rho_0^2+\omega_0^2}{\omega_0^2}\left[\left(\frac{\hbar^2\omega_0^2m}{k_BT'}+i\frac{\hbar^2\rho_0\omega_0m}{k_BT'}\right)e^{(-\rho_0-i\omega_0)\tau}+c.c.\right]~.
\end{split}
\end{equation}
Hence the perturbed noise spectrum can be obtained by integrating  $\delta\left\langle x(\tau) x(0)\right\rangle_{H,R} e^{i\omega\tau}$over all possible values of $\tau$ . We keep in mind that the entire expression is valid for $\tau\geq0$ and in the regime $\tau< 0$ it has no contributions. The result obtained is as follows
\begin{equation}\label{A5.16}
\begin{split}
\delta S(\omega)&=\int_{-\infty}^\infty \delta\left\langle x(\tau) x(0)\right\rangle_{H,R} e^{i\omega\tau}d\tau\\
&=\int_0^\infty \delta\left\langle x(\tau) x(0)\right\rangle_{H,R} e^{i\omega\tau}d\tau\\
&\cong\frac{24\hbar^2\mathcal{A}^2G^2k_BT'(\alpha^2+2\gamma)\omega^2(\omega^2-\Omega^2)}{m\kappa(\rho^2\omega^2+(\omega^2-\Omega^2)^2)^2}+\frac{3\hbar^4\mathcal{A}^2G^2k_BT'(4\gamma-13\alpha^2)\Omega^2}{2k_BT'm\kappa(\rho^2\omega^2+(\omega^2-\Omega^2)^2)}\\
&\cong \frac{12k_B^2T'^2(\alpha^2+2\gamma)\rho\omega^2(\omega^2-\Omega^2)}{(\rho^2\omega^2+(\omega^2-\Omega^2)^2)^2}+\frac{3(4\gamma-13\alpha^2)\rho\Omega^2\hbar^2}{4(\rho^2\omega^2+(\omega^2-\Omega^2)^2)}~.
\end{split}
\end{equation}
In the free mass limit eq.(\ref{A5.16}) (i.e. $\omega\gg\Omega$) takes the form given as
\begin{equation}\label{A5.17}
\delta S(\omega)\simeq\frac{3\rho\hbar^2}{\omega^2}\frac{\Omega^2}{\omega^2}\left\{8\left(\frac{\alpha^2}{2}+\gamma\right)\left(\frac{k_BT'}{\hbar\Omega}\right)^2+\left(\gamma-\frac{13\alpha^2}{4}\right)\right\}~.
\end{equation}
At resonance frequency (i.e. $\omega=\Omega$) eq.(\ref{A5.16}) takes the form
\begin{equation}\label{A5.18}
\delta S(\Omega)=3\left(\gamma-\frac{13\alpha^2}{4}\right)\frac{\hbar^2}{\rho}~.
\end{equation}

\end{document}